\documentclass[pdflatex,sn-mathphys-num]{sn-jnl}

\usepackage{graphicx}%
\usepackage{multirow}%
\usepackage{amsmath,amssymb,amsfonts}%
\usepackage{amsthm}%
\usepackage{mathrsfs}%
\usepackage[title]{appendix}%
\usepackage{xcolor}%
\usepackage{textcomp}%
\usepackage{manyfoot}%
\usepackage{booktabs}%
\usepackage{algorithm}%
\usepackage{algorithmicx}%
\usepackage{algpseudocode}%
\usepackage{listings}%

\theoremstyle{thmstyleone}%
\theoremstyle{thmstyletwo}%

\theoremstyle{thmstylethree}%

\begin{document}

\title[Article Title]{Legal Entanglement}


\author[1]{\fnm{Nicholas} \sur{Godfrey}}
\author[2]{\fnm{Ted} \sur{Sichelman}}
\email{nic.godfrey@qut.edu.au} 
\email{tsichelman@sandiego.edu}
\equalcont{These authors contributed equally to this work.}

\affil[1]{\orgdiv{School of Law}, \orgname{Queensland University of Technology}, \orgaddress{\street{2 George Street}, \city{Brisbane}, \postcode{4011}, \state{Queensland}, \country{Australia}}}

\affil[2]{\orgdiv{Law School}, \orgname{University of San Diego}, \orgaddress{\street{5998 Alcala Park}, \city{San Diego}, \postcode{92130}, \state{CA}, \country{USA}}}


\abstract{Quantum entanglement is a phenomenon in which two physical systems are correlated in such a way that they appear to instantaneously affect one another, regardless of the distance between them. As commonly understood, Bell’s Theorem famously demonstrates that any causal explanation of entanglement must discard either locality (the principle that nothing, including information, travels faster than light) or classical notions of realism (or both). Drawing on this concept, several legal scholars have metaphorically described 'entangled' legal concepts. For instance, if a state’s highest court redefines the concept of 'foreseeability' in negligence law, this redefinition alters the concept of 'reasonable care' immediately in the eyes of the law. Godfrey (2024) is the first work to mathematically model entangled legal concepts, particularly in the context of legal interpretation. Here, we extend the quantification to the formulation and delineation of law (lawmaking) and the adjudication of law (judgment). In so doing, we connect legal entanglement to Sichelman’s (2022) work on legal entropy, complexity, and the informational content of law. In addition to quantifying entanglement across various legal contexts, our approach provides broader insights. For example, it offers a more comprehensive analysis of the uses and limits of 'modularity' in law—specifically, the role legal boundaries (spatial or intangible) play in reducing information costs within legal systems. Moreover, we discuss how our model can improve theories of legal artificial intelligence. Finally, we explore the application of legal theory back to physics. If quantum physical entanglement operates analogously to legal entanglement, it requires discarding both locality and classical realism, though not in the manner commonly imagined.}

\keywords{legal entanglement, quantum entanglement, John Bell, legal interpretation, legal formulation, adjudication, indeterminacy, entropy, uncertainty, complexity, information theory, modularity, nonlocality, causality, realism, Hohfeld, quantum-like}



\maketitle

\section{Introduction}\label{sec1}

John Bell, perhaps the most famous of all interpreters of quantum mechanics, once remarked, '[T]here \textit{are} things which \textit{do} go faster than light . . . . When the Queen dies in London . . . the Prince of Wales, lecturing on modern architecture in Australia, becomes \textit{instantaneously} King' \cite{bell2004} (Bell, Speakable and Unspeakable, 2nd ed., 2004 p. 234). 

Of course, the instantaneous transition of the Prince to King is a legal process; importantly, one that is amenable to legal analysis, particularly under the scheme introduced by Wesley Hohfeld \cite{hohfeld1913} (1913) at the turn of the 20th century. According to Hohfeld, when the Queen dies, a constellation of 'regal' jural relations held by the Queen (powers, rights, privileges, and the like) are extinguished, and instantaneously (at least in an ideal sense) those regal relations become those of the former Prince, now King. 

More generally, when a jural relation changes -- whether it be the death of a monarch or the issuance of a judicial opinion -- jural relations 'entangled' with the changed legal relation generally will also instantaneously (again, an idealized sense) and nonlocally change. For instance, when a nation's highest court reinterprets a statutory term that is used throughout a legal code, that court exercises a jural power that instantaneously updates the meaning of that term across the statutory code, which in turn will typically instantaneously update the legal relations among legal actors that involve the relevant provisions of the code. Suppose, for example, the highest court redefines 'foreseeability' so that it is more difficult for defendants to show that a certain type of harm was not foreseeable. Assuming the interpretation applies retroactively, the reinterpretation will instantaneously make certain would-be defendants liable for negligence who were not liable before the judicial decision. 

In this paper, we posit that the instantaneous updating of legal relations following a judicial decision (adjudication) is a form of 'legal entanglement' in the sense that a legal action changing a given 'local' legal relation instantaneously and nonlocally updates non-local legal relations 'entangled' with that local relation.\footnote{At this point in the paper, we use the terms 'jural relation' and 'legal relation' interchangeably.} So far, there is nothing particularly 'quantum' about such instantaneous and nonlocal legal phenomena. More specifically, again using adjudication for explication, if the outcome of the judicial decision is perfectly predictable (at least given all the relevant facts and law for the decision), the decision is \textit{classical} in the Hohfeldian sense and the instantaneous and nonlocal updating of 'entangled' relations is also classical. The fact that \textit{physical} laws do \textit{not} admit of such 'classical', non-local entanglement, of course, does not preclude such phenomena from appearing in legal and other contexts. \footnote{In both physics and data science, 'classical entanglement' is generally used to refer to certain \textit{local} correlations that otherwise resemble quantum correlations. However, in law, following the convention of Hohfeld and most legal theorists, 'classical' legal phenomena simply refers to deterministic phenomena, whether local or nonlocal. Hence, from the legal perspective, nonlocal updating of 'entangled' legal relations may indeed be classical in the Hohfeldian sense.}

On the other hand, since at least the advent of the Legal Realist movement, legal theorists have posited that at least some legal relations -- at least prior to a legal judgment -- are inherently indeterminate in the sense that they cannot be precisely specified, even by experts who are aware of all the relevant law and facts. On such a post-classical view, at least in 'hard' cases, legal relations exist in a 'superposition' to the degree that the best available description of the relations cannot distinguish between multiple possible states that are possible results of judgment \cite{sichelman2024} (Sichelman, 2025a). Instead, like quantum mechanical states, the legal states can only (if even then) be assigned probabilities of being 'measured' upon judgment. This indeterminacy is for the realists not merely epistemological but wholly ontological in nature, because even in an idealized world in which all the relevant facts and law are available to expert legal observers, the nature of the legal relation can still not be known. Indeed, one can even question if the notion of a 'determinate' relation is sensible prior to judgment.

For example, judicial decisions in 'close' patent infringement actions are often notoriously difficult to predict, even in an idealized world with all of the relevant facts and law in hand. In a difficult case, experts can only assign probabilities, such as 60 percent that the patent is infringed and 40 percent that the patent is not, which materialize into 'classical' legal states (in the Hohfeldian sense) upon judgment. In other words, it is only upon judgment that the patent is either infringed (100 percent) or not (0 percent). The pre-judgment legal statement is metaphorically akin to a 'quantum superposition' of physical states and the judgment is akin to a 'quantum measurement', which collapses the superposition to a classical-like eigenstate \cite{sichelman2024} (Sichelman, 2025a). 

With this notion of quantum-like legal relations in hand, we can introduce a notion of 'quantum legal entanglement'. Specifically, quantum legal entanglement occurs when legal states exist in a superposition prior to some form of legal 'measurement' and the change in those legal states entails instantaneous and nonlocal changes in other 'entangled' legal states. These changes further entail that measurement of these other entangled legal states will yield different outcomes than those that would have occurred absent entanglement. In what follows, we unpack this notion in three key legal contexts: the interpretation of the law (particularly by the courts); the formulation and delineation of the law (often by legislatures but sometimes by courts); and the adjudication of legal disputes (generally by courts, but also by governmental agencies, private arbitrators, and other governing bodies).

This paper is a notable contribution to the literature in four respects. First, although previous scholars have recognized that physical entanglement can be used to metaphorically describe certain aspects of the law, we provide a richer conceptual account of how quantum entanglement can play a role in legal systems. Second, extending our prior work (especially \cite{godfrey2024} Godfrey (2024)), we offer a detailed quantitative formalism to characterise the role of entanglement across multiple legal domains. Third, leveraging this formalism, we discuss important theoretical and practical applications of our formalism, including the interpretation, formulation, and adjudication of the law as well as the role of information costs and 'modularity' in the law and legal artificial intelligence. Finally, we apply our conceptualisation of legal entanglement in reverse to physical entanglement to offer an original proposal for understanding the nature of quantum entanglement.

More specifically, in Section 2, we provide a brief introduction to quantum entanglement in physics and data science. In Section 3, incorporating and adapting the mathematical framework of physical entanglement, we define and describe legal entanglement in the contexts noted above. We briefly remark upon the theoretical and practical applications of legal entanglement, both within and outside the legal context. In the legal context, our formalism can aid in developing richer AI legal systems. Outside of the law, if we assume that physical entanglement functions similarly to legal entanglement, our formalism offers a novel approach to understanding quantum entanglement. More precisely, this account supports a non-local, non-realistic approach, though one with quite different ramifications for jettisoning locality and realism than commonly believed.

\section{Quantum Entanglement}\label{sec2}

In the world of classical information theory, we typically use the \textit{bit}, a unit of information that can exist in one of two states: $0$ or $1$. For example, whether a light bulb is on or off can be represented by a single bit. In order to shift to quantum information theory, we trade the bit for the simplest non-trivial quantum system, the \textit{qubit}, which exists in a 'superposition' of 'orthonormal' \textit{basis states}, $|0\rangle$ and $|1\rangle$. Unlike the light bulb, the qubit represents a system that is neither on nor off. Rather, upon a so-called measurement, the qubit collapses to one of these two states.\footnote{The situation is even more complicated, because in quantum mechanics, the basis states will typically change depending upon the nature of the measurement being made. For instance, one can measure an electron's quantum spin along different axes, with each axis associated with different basis states. For simplicity, we assume for now that there is only one basis of interest.} 

A general state of a single qubit, $|\psi\rangle$, can be expressed as a linear combination of its basis states:

\begin{align}
|\psi\rangle = \alpha|0\rangle + \beta|1\rangle,
\label{basicQubitWavefunction}
\end{align} 
where $\alpha$ and $\beta$ are complex numbers which are used to determine the probability of the qubit being measured in either the $|0\rangle$ or $|1\rangle$ states. Specifically, the probability of $|\psi\rangle$ collapsing to $|0\rangle$ (or $|1\rangle$) is $|\alpha|^2$ (or $|\beta|^2$). Because the qubit must collapse to either $|0\rangle$ or $|1\rangle$, the probabilities of each state being measured must add to 100 percent, such that $|\alpha|^2+|\beta|^2=1$. 

Notably, under the standard interpretations of quantum mechanics, the superposition of basis states does not reflect a lack of knowledge by observers of the actual state of the system, as in classical statistical mechanics. Rather, the indeterminacy is considered ontological in nature, such that until measurement, the system exists in both states (or perhaps in neither state, depending upon one's take). Either way, upon the standard interpretation, there is no definite state of the quantum system modeled by two or more basis states prior to measurement. As noted in Section 1, this aspect of quantum states is central to the analogy between quantum and nondeterministic legal states.

Mathematically, the quantum state $|\psi\rangle$ may be modeled as a complex vector that exists in a vector space known as a \textit{Hilbert space}. The Hilbert space of a qubit is two-dimensional, as it can be measured in one of two states. By definition, the dimension of a Hilbert space for a system is equal to the number of orthonormal basis states -- essentially the number of outcomes upon measurement -- that can be defined in that system. 

However, we can also model higher-dimensional systems. For instance, the qutrit has three orthonormal basis states. And so on. In general, the Hilbert space dimension of a system modeled by qubits will be $2^n$, where $n$ is the number of qubits needed to model the system. For example, a two-qubit system can be represented by vectors in a four-dimensional Hilbert space:
\begin{align}
    |\psi\rangle = \alpha |00\rangle + \beta |01\rangle + \gamma |10\rangle + \delta |11\rangle, \label{2qubitWF}
\end{align}
where a two-digit basis state is used to represent the `joint' state space of two single-qubit basis states via the tensor product:\footnote{Here, $\otimes$ denotes the tensor product.}
\begin{align}
    |ij\rangle \equiv |i\rangle \otimes |j\rangle.
\end{align}

For instance, the `joint' state space may represent the spins of two different particles, A and B, where particle A may be measured as $|0\rangle$ or $|1\rangle$ and particle B may be measured as $|0\rangle$ or $|1\rangle$. If all measurement possibilities are allowed, particle A could be measured as $|0\rangle$ and particle B as $|0\rangle$, particle A as $|0\rangle$ and particle B as $|1\rangle$, and so forth, resulting in four different possible outcomes, $|00\rangle$,  $|01\rangle$, $|10\rangle$, and $|11\rangle$.

The state described by Equation \eqref{2qubitWF} is an example of a \textit{bipartite} state, as it has two degrees of freedom, each corresponding to one qubit that can be measured in one of two states, here, $|0\rangle$ or $|1\rangle$. In some cases, we can write a bipartite state as the tensor product of two single-qubit states. For example, the bipartite state
\begin{align}
    |\psi\rangle = \alpha_0 \alpha_1 |00\rangle + \alpha_0 \beta_1 |01\rangle + \beta_0 \alpha_1 |10\rangle + \beta_0 \beta_1 |11\rangle, \label{separableSystem}
\end{align}
can be factorised and rewritten as
\begin{align}
|\psi\rangle &= \left(\alpha_0 |0\rangle + \beta_0 |1\rangle\right) \otimes \left(\alpha_1 |0\rangle + \beta_1 |1\rangle\right) \nonumber \\ &=    |\psi_1\rangle \otimes |\psi_2\rangle,
\end{align}
where $|\psi_i\rangle = \alpha_i |0\rangle + \beta_i |1\rangle$.

Such states are known as \textit{separable states}. A state that is not separable is an \textit{entangled} state. In other words, if two systems are entangled, their states cannot be described independently of each other. Mathematically, we can see this by considering the state,
\begin{align}
    |\Phi^+\rangle = \frac{1}{\sqrt{2}} \left(|00\rangle + |11\rangle\right),
\end{align}
which is one of four possible maximally entangled two-qubit states, also known as Bell states. It is trivial to see that such a state cannot be decomposed into the tensor product of two separate systems:
\begin{align}
    \nexists \; \alpha_0, \alpha_1, \beta_0, \beta_1 : |\Phi^+\rangle = \left(\alpha_0 |0\rangle + \beta_0 |1\rangle\right) \otimes \left(\alpha_1 |0\rangle + \beta_1 |1\rangle\right).
\end{align}
In other words, the two states cannot act independently of one another. For the entangled state $|\Phi^+\rangle$, measuring one particle in the state $|j\rangle$ would immediately require the other particle to be measured in state $|j\rangle$. Specifically, if one particle is measured in the $|0\rangle$ state, so will the other, and similarly if one particle is instead measured in the $|1\rangle$ state (nor does it matter which particle is measured first).

Physically, quantum entanglement has profound implications that challenge the intuition offered by classical mechanics. Specifically, under plausible assumptions, entanglement requires us to abandon the notion of a locally real universe.\footnote{One can save locality and realism with implausible assumptions, such as 'superdeterminism', in which the entire evolution of the universe, including all human behavior, is fully determined, but here we adhere to the plausible assumptions of the standard approach to entanglement.} In other words, entanglement is not compatible with a universe that maintains both locality -- the notion that nothing travels faster than the speed of light -- and realism -- the notion that all measurable properties of physical systems have fixed values prior to measurement. 

While the full scope and import of quantum entanglement could occupy multiple volumes, in the next section, we focus on the implications of the conceptual and mathematical structure of entanglement for legal systems, though we briefly return to notions of physical locality and realism at the end.

\section{Entanglement in the Law}\label{sec3}

To be certain, previous scholars have posited in a purely conceptual sense that the physical notion of 'entanglement' plays a role in legal systems (\cite{feldman2012}Feldman, 2012; \cite{krisch2021}Krisch, 2021). For example, Feldman  \cite{feldman2012}(2012) explains that '[i]n entangled [legal] concepts, the descriptive and the evaluative are fundamentally interrelated such that when one aspect is reshaped so is the other.'

Providing a conceptualization of law using physical concepts can be useful as a first step in achieving practical and theoretical benefits, but fully achieving that goal in our view requires the incorporation and adaptation of the relevant mathematical framework underlying those physical concepts. In this respect, our prior work \cite{godfrey2024} (Godfrey (2024)) and this paper are the first to utilise the mathematical description of entanglement from physical systems to offer a quantitative account of the entanglement of legal systems. In so doing, we also offer a richer conceptual account of entanglement than previous scholars.

Such an extension might seem extreme at first -- the physical pages or screens on which we read and write laws are macroscopic and, hence, do not display quantum properties. Indeed, we do not propose that legal rules \textit{physically} exhibit quantum properties. However, we have previously shown that the mathematical tools of quantum theory can fruitfully be applied to model bodies of legal rules and relations. These models turn on the quantum-like nature of nondeterministic legal rules and relations that exist in legal 'superpositions' described in Section 1. (For further discussion, see \cite{sichelman2024} Sichelman (2025a) and \cite{godfrey2024} Godfrey (2024).)

Furthermore, depending on the legal ontology one constructs, we propose that the rejection of local realism demanded by physical entanglement extends from the quantum world to the legal world. Locality, for example, is violated \textit{in theory} in the legal ontology by judgments and the enactment of new laws. Of course, physically, the information contained in a judgment or statute cannot travel faster than the speed of light, and hence cannot violate physical locality. However, in an ideal \textit{legal} world, the passing of a bill or judicial interpretation of a legal rule updates the law immediately, hence implying 'non-locality' in a purely legal ontology, even if this non-locality is not \textit{physically} realisable.

Non-realism is perhaps more controversial, and dependent upon the reader's jurisprudential leanings. Again, as discussed more comprehensively in our previous works,\cite{sichelman2024, godfrey2024} different theories of law vary according to the allowed sources of legal indeterminacy. Adoption of Ronald Dworkin's \cite{dworkin1977} (1977) 'right answer' thesis, for example, would imply that all legal indeterminacy is merely epistemic -- there always exists a correct meaning or application of a legal rule, and any indeterminacy regarding that rule is due solely to a lack of information. 

In stark contrast, as noted earlier, the Legal Realists advocated that indeterminacy is inherent to the law, and that judges, in interpreting and applying legal rules, do not merely discover the law, but make it. From this perspective, the indeterminacy of the law is not of an epistemic nature, but rather ontological. Under Legal Realism, there is no 'correct' interpretation of a given rule until it has been decided by the judiciary, in much the same way that the properties of quantum systems cannot be fixed prior to measurement. Somewhat confusingly, the position of Legal Realism on legal indeterminacy can be equated to the non-realist, purely contextual description of physical reality -- there is no classical-like property of a system (such as position or momentum in physics, or liability or no liability in law) until it is measured and effectively 'realised'.

Regardless of whether one wishes to go so far as to reject the legal equivalents of both locality and realism in legal systems, so long as one rejects either element, the ontology of law can be considered at least quantum-like, enough so to prohibit a locally real universe. Following this line of thinking, in this section, we apply the mathematical formalism of quantum information to consider various forms of quantum entanglement in the law.\footnote{As we noted in Section 1, legal systems also admit of a form of classical, non-local entanglement, though we leave further discussion of this notion to future work.}

\subsection{Interpretive}\label{sec:Interpretive}

\subsubsection{Internal Entanglement}

The basis states, such as the $|0\rangle$ and $|1\rangle$ states used in Equation \eqref{basicQubitWavefunction}, are typically used to represent orthogonal characteristics of a system -- in other words, different possible outcomes of a measurement. The basis states might represent physical properties of particles, like spin-up or spin-down:
\begin{align}
    |\psi\rangle = \alpha |\uparrow\rangle + \beta |\downarrow\rangle.
\end{align}
However, they can also represent any other orthogonal characteristics. Take, for instance, Schr\"{o}dinger's infamous cat, which may be observed as dead or alive:
\begin{align}
    |\text{Cat}\rangle = \alpha |\text{Dead}\rangle + \beta |\text{Alive}\rangle.
\end{align}
In this section, we outline the representation of orthogonal legal rule interpretations as qubit basis states. That is, we model an ambiguous legal rule with two interpretations, $I_1$ and $I_2$, as:
\begin{align} 
    |\psi\rangle = \alpha |I_1\rangle + \beta |I_2\rangle.
\end{align}
For example, Section 1023D of Australia's \textit{Corporations Act 2001} (Cth) refers to 'significant detriment' to consumers, which either has been, or is likely to be, caused by financial products. Suppose Alice is unsure whether a non-monetary detriment (e.g., mental anguish) could satisfy Section 1023D's 'significant detriment' requirement, or whether the detriment is limited to monetary loss. Alice's uncertainty gives rise to two orthogonal interpretations,
\begin{align}
    |\text{Detriment}\rangle_\text{Monetary} = m |\text{Strictly Monetary}\rangle + \bar m |\neg\text{Strictly Monetary}\rangle,
\end{align}
where $|m|^2$ ($|\bar m|^2$) is the probability that significant detriment is (is not) limited to strictly monetary detriment. 

Like general quantum systems, we can also consider higher-dimensional interpretations that combine multiple levels of interpretive indeterminacy. For example, independent of the indeterminacy surrounding the necessity of a monetary nature of 'significant detriment', Alice might also wonder whether 'significant detriment' can be met merely by harm to a single specific consumer or whether it must affect a wide class of consumers:
\begin{align}
    |\text{Detriment}\rangle_\text{Widespread} = w |\text{Widespread}\rangle + \bar w |\neg\text{Widespread}\rangle.
\end{align}
If we now wish to represent the interpretation of 'significant detriment' as a whole using qubits, we must use a Hilbert space of dimension 4, because there are two possibilities along the 'monetary' dimension and two possibilities along the 'widespread' dimension:
\begin{align}
        |\text{Detriment}\rangle &= |\text{Detriment}\rangle_\text{Monetary} \otimes |\text{Detriment}\rangle_\text{Widespread} \nonumber\\
        &= mw|MW\rangle + \bar m w|(\neg M) W\rangle + m \bar w |M(\neg W)\rangle + \bar m \bar w |(\neg M ) (\neg W)\rangle.
        \label{1023D_Detriment}
\end{align}
So far, the indeterminacy in these examples could easily be represented by classical probability theory. However, as with physical systems, we can entangle interpretive components to represent inherent relationships within an interpretation that are more easily modeled with quantum mechanical approaches. Consider, for example, Section 301D of the \textit{National Consumer Credit Protection Act 2009} (Cth), which was implemented by the same amendment that introduced our earlier provision, Section 1023D of the \textit{Corporations Act}, and which shares much of the same wording as  Section 1023D, including reference to 'significant detriment'. 

Given the common origin of the two terms, legal observers would generally assume that the legislature's intended definition of 'significant detriment' for the purposes of Sections 1023D and 301D ought to be identical. In other words, suppose we have a final judgment from the High Court of Australia interpreting 'significant detriment' in Section 1023D, which is in turn binding on all other courts. That is, if, upon judicial measurement, one finds $|\text{Detriment, 1023D}\rangle$ in state $|AB\rangle$, where $|AB\rangle$ is one of the four basis states in Equation \eqref{1023D_Detriment}, then the interpretation of $|\text{Detriment, 301D}\rangle$ ought to immediately become $|AB\rangle$ as well. We can achieve this by applying a set of quantum gates, $U$, such that the Section 1023D and Section 301D interpretations are entangled:
\begin{align}
    |\text{Detriment}\rangle &= U\left(|\text{Detriment, 1023D}\rangle \otimes |\text{Detriment, 301D}\rangle\right) \nonumber \\
    &=  \alpha (|MW\rangle_{\text{1023D}}\otimes|MW\rangle_{\text{301D}}) + \beta (|M \bar{W}\rangle_{\text{1023D}} \otimes |M \bar{W}\rangle_{\text{301D}}) \nonumber \\ &\quad + \gamma (|\bar{M} {W}\rangle_{\text{1023D}} \otimes |\bar{M} {W}\rangle_{\text{301D}}) + \delta (|\bar{M} \bar{W}\rangle_{\text{1023D}} \otimes |\bar{M} \bar{W}\rangle_{\text{301D}}) \nonumber \\
    &=  \alpha |0000\rangle + \beta |0101\rangle + \gamma |1010\rangle + \delta |1111\rangle, 
    \label{eq:DetrimentState}
\end{align}
where $\alpha,\beta,\gamma,\delta$ are complex probability amplitudes whose absolute squares are the probabilities of the associated interpretation being measured (i.e., interpreted by the court) and, in the final line, we substitute $M,W \Rightarrow 0$ and $\bar{M}, \bar{W} \Rightarrow 1$ (in other words, we just substitute numbers for letters).  After applying the quantum gates $U$, there are four unique outcomes, wherein the measured state of the first and second qubit (representing Section 1023D's interpretations) always matches the measured state of the third and fourth qubit, respectively (representing Section 301D's interpretive possibilities). In essence, we have linked the two interpretations so that judgment with respect to one will instantaneously and non-locally update the interpretation of the other, akin to the measurement of two maximally entangled quantum states.

\subsubsection{External Entanglement}

Thus far, we have considered the ambiguity that is 'internal', or 'baked-in', to a legal rule -- ambiguity that stems from the meaning of the rule itself. But we may also want to consider a higher level of indeterminacy in a law's interpretation. For example, could we distinguish between a rule's internal ambiguity, and its additional ambiguity stemming from its dependence on another ambiguous rule's interpretation? Or could we consider two competing representations of an ambiguous rule -- both of which acknowledge ambiguity within a rule, but disagree on how to assign probability amplitudes to the rule's various interpretations? Indeed, we can represent both sets of these 'external' ambiguities by borrowing another tool from quantum theory -- mixed states.

In the language of quantum theory, the objects we have used thus far, like $|\text{Detriment}\rangle$, are known as 'pure states'. A probabilistic mixture of multiple pure states is known as a mixed state, which is typically represented as a sum of pure state matrices, each multiplied by the probability corresponding to that pure state, $p_i$:\footnote{Readers will note that our notation has changed here, as we now treat states as matrices ($|\psi\rangle \langle \psi|$) rather than vectors ($|\psi\rangle$). This change is necessary to represent density matrices, but for brevity's sake, we will not dwell on why. A more in-depth exploration of the mathematical structure of pure and mixed interpretive states in the context of legal states is provided in \cite{godfrey2024} Godfrey (2024).}

\begin{align}
    \rho = \sum_i p_i |\psi_i\rangle\langle \psi_i|.
\end{align}

We can use these 'density matrices' to represent rules with both internal and external ambiguities. For example, suppose law students, Alice and Bob, are both asked to predict a court's interpretation of a rule, $R$, with possible interpretations $|0\rangle$ and $|1\rangle$. They return with two competing predictions:
\begin{align}
    |R\rangle_A = \sqrt{0.2} |0\rangle + \sqrt{0.8}|1\rangle, \quad |R\rangle_B = \sqrt{0.7} |0\rangle + \sqrt{0.3}|1\rangle.
\end{align}
Based on past performance, perhaps we believe Alice is $90\%$ likely to have accurately modelled the rule, versus a $10\%$ likelihood that Bob's model is in fact correct. Accordingly, we set $p_A = 0.9$, $p_B = 0.1$. Our mixed state, which now incorporates both the rule's inherent ambiguity as well as the 'external' uncertainty regarding that ambiguity, is represented as
\begin{align}
    \rho = 0.9 |R\rangle_A\langle R|_A + 0.1 |R\rangle_B\langle R|_B.
\end{align}

Consider also a scenario in which one rule's interpretation, $R_A$, is entangled with another's interpretation, $R_B$ -- just as we argued Sections 1023D and 301D's interpretations were entangled -- and we wish to represent $R_A$ in isolation. In physical quantum systems, we can achieve this by 'tracing out' the unwanted rule. In quantum mechanics, this is achieved by applying an operation known as a partial trace over subsystem $B$ to the density matrix: $\rho_A =\text{Tr}_B \rho$. We can do the same operation on our entangled legal interpretations. 

Suppose, for example, we are maximally uncertain as to how to interpret rules $A$ and $B$, but we are certain that if one is interpreted as $|1\rangle$, the other must be too:
\begin{align}
    |\psi\rangle = \frac{1}{\sqrt{2}} \left(|0\rangle_A \otimes |0\rangle_B + |1\rangle_A \otimes |1\rangle_B\right)
\end{align}
In practical statutory interpretation terms, this might represent two rules in the same provision that both use the same ambiguous word. To represent rule $A$ independently, we first represent both rules as a density matrix, $\rho_{AB}$:
\begin{align}
    \rho_{AB} = |\psi\rangle\langle\psi|.
\end{align}
Taking the partial trace over $B$ of $\rho_{AB}$ leaves us with
\begin{align}
    \rho_A =\text{Tr}_B \rho_{AB} = \frac{1}{2}(|0\rangle\langle0| + |1\rangle\langle1|).
\end{align}
This mixed state, which represents rule $A$, contains external ambiguity as opposed to the internal ambiguity present in pure states. This tells us that our uncertainty surrounding rule $A$ arises not from within the rule itself, but from some external source of ambiguity. 

These various representations of legal interpretations allow us to better characterise statutory indeterminacy.\footnote{A more in-depth consideration of these measures is provided in (Godfrey, 2024).} Furthermore, as we touch upon in Section 3.4, we can apply analytical measures from the quantum toolbox to unlock new insights into the relationships between various interpretations. In the meantime, we examine legal entanglement in the context of the formulation and adjudication of the law.

\subsection{Formulative}\label{sec:Formulative}

The formulation and delineation of law by legislatures and government agencies -- typically in the form of statutes and regulations -- is an exercise of legal power to change the law \cite{sichelman2024}(Sichelman, 2025a). For instance, when the U.S. Congress passed the America Invents Act in 2011 (AIA), it significantly expanded the ability of would-be patent infringers and other parties to challenge the validity of already issued patents in the U.S. Patent \& Trademark Office (USPTO).  These new procedures did not per se change the law of patent validity. Nonetheless, they significantly increased the likelihood that a given patent would be found invalid, because they reduced the pre-litigation probability that the typical patent could be effectively asserted against a would-be infringer. In effect, a change in procedure instantaneously altered the substantive law, to the extent that substantive law should be viewed simply as 'prophecies of what the courts will do in fact' \cite{holmes1897}(Holmes, 1897).

In other words, a legislative enactment that directly modifies one area of the law may instantaneously affect -- often, very broadly -- legal relations that are 'nonlocally' entangled with the area of the law being altered. In contrast to quantum mechanics -- a system where the 'global' laws are 'fixed' -- because the legal system does allow for global changes in law, instantaneous effects appear even before the remote legal state is measured, in essence, rotating the state vector of the remote state to another probabilistic, pre-measurement state. We can model such changes via a quantum operator, \textit{P}, such that the operator's effect on legal state \textit{A} in turn instantaneously changes a remotely entangled state \textit{B}.  

To model this relationship, assume two legal subsystems \( A \) and \( B \) are in an entangled legal state:
\[
|\Psi_{AB}\rangle = \sum_{i} \alpha_i \, |a_i\rangle \otimes |b_i\rangle
\]

A legislative operator \( \mathcal{P}_A \) acts on subsystem \( A \), such that:
\[
\mathcal{P}_A \otimes \mathbb{I}_B \, |\Psi_{AB}\rangle = \sum_i \alpha_i \, \mathcal{P}_A |a_i\rangle \otimes |b_i\rangle
\]

Although \( \mathcal{P}_A \) acts locally, the entanglement causes an instantaneous transformation of the joint state, thereby altering the marginal state of \( B \), as illustrated by tracing out the state of \( A \) to obtain the reduced density matrix of \( B \), 
\[
\rho_B = \mathrm{Tr}_A \left( \mathcal{P}_A \otimes \mathbb{I}_B \, |\Psi_{AB}\rangle \langle \Psi_{AB}| \, \mathcal{P}_A^\dagger \otimes \mathbb{I}_B \right).
\]

Thus, even without measurement, the legal subsystem \( B \) is probabilistically 'rotated' due to the transformation of its entangled partner \( A \). This rotation in effect alters the probability that some related provision in a statute that was not directly changed by the legislature will be interpreted in a particular fashion. 

That the operator \textit{P} affects the remote state prior to measurement is critical to understanding the decisions legal actors make when the scope or application of laws cannot be predicted with certainty.  In quantum mechanics,  the reduced state of a subsystem \( B \) that is entangled with a remote subsystem \( A \) (and is otherwise immune to outside influences) remains invariant prior to measurement, because other than measurement, any operation applied to \( A \) will be unitary and cannot affect subsystem \( B \).

More precisely, in quantum mechanics, if \( |\Psi_{AB}\rangle \) is a pure entangled state and \( U_A \) is a unitary operator acting only on subsystem \( A \), then:

\begin{equation}
\rho_B = \mathrm{Tr}_A \left[ (U_A \otimes \mathbb{I}_B) |\Psi_{AB}\rangle \langle \Psi_{AB}| (U_A^\dagger \otimes \mathbb{I}_B) \right] = \mathrm{Tr}_A \left[ |\Psi_{AB}\rangle \langle \Psi_{AB}| \right]
\end{equation}

This invariance of \( \rho_B \) reflects the restrictions of the \emph{no-signaling theorem} in quantum theory: local unitary operations on subsystem \( A \) cannot instantaneously affect the observable statistics of subsystem \( B \) in the absence of measurement or classical communication. Consequently, standard quantum mechanics prohibits the transmission of information faster than light via entanglement alone.

In contrast, under our proposed legal analogy, a legislative enactment represented by an operator \( \mathcal{P}_A \) applied to a legal subsystem \( A \) may instantaneously modify a nonlocally entangled subsystem \( B \), even prior to any 'measurement' (i.e., judgment) in \( B \). That is, the operator \( \mathcal{P}_A \) transforms the joint legal state in a way that results in a new reduced state \( \rho_B' \neq \rho_B \), despite no direct action having occurred in subsystem \( B \). 

This divergence from standard quantum behavior suggests that certain legal transformations -- here, legislation -- are not analogous to unitary operators in Hilbert space, but rather to non-unitary, globally state-altering superoperators or rule-updating transformations. These legal operators may redefine the space of permissible states or restructure the form of entanglement itself, thereby enacting a second-order modification to underlying legal relations -- in Hohfeldian terms, the exercise of a 'global' legal power \cite{hohfeld1913} (Hohfeld, 1913).  Accordingly, while unitary dynamics preserve local marginal states, legal enactments may globally rotate or reproject the legal state vector in ways that do not respect the constraints of no-signaling. 

Yet, quantum measurement is also a non-unitary process, and indeed, quantum entanglement in the physical sense is only manifested following this non-unitary process. As we discuss further in Section 3.5, if we reverse the legal analogy and apply it to the physical setting, then the non-unitary process of measurement must, at least upon the standard Copenhagen interpretation, be understood not merely as an epistemic update, but as an ontology-altering, second-order operation akin to legal judgment \cite{sichelman2024} (Sichelman, 2025a). This interpretation implies that quantum entanglement reflects both a contextual and non-local physical process. 

\subsection{Adjudicative}\label{sec:Adjudicative}

Unlike formulative entanglement, which results in the instantaneous updating of the state of entangled systems even absent a measurement, adjudicative entanglement is manifested only upon a legal measurement, namely, a legal judgment (i.e., adjudication) by a court. As noted earlier, the legal state of a dispute in court can be readily modeled as a superposition between two outcomes:
\begin{align} 
    |\psi\rangle = \alpha |\text{Liability}\rangle + \beta |\text{No-liability}\rangle.
\end{align}
Like a court deciding between one legal interpretation and another, in adjudicative processes, courts will ultimately determine whether a defendant is liable or not with respect to a particular claim brought by the plaintiff. If we adhere to the post-classical approach, prior to judgment, a legal state will exhibit ontological indeterminacy in the sense that even all of the applicable law and facts will be insufficient to determine the outcome of adjudication. In this sense, the legal state is inherently indeterministic and akin to a probabilistic quantum superposition (\cite{sichelman2024} Sichelman, 2025a; \cite{godfrey2024} Godfrey, 2024).

As described earlier, the essence of quantum entanglement is that measurements performed on one part of an entangled system are correlated with those on the other part in a way that cannot, in general, be accounted for by any classical theory based on locally realistic hidden variables. In other words, quantum entanglement cannot be explained by measurement outcomes that are pre-determined by local variables acting only within their immediate surroundings. 

Our claim here is that the adjudicatory process exhibits a similar form of entanglement that cannot be explained by hidden, locally real variables. For instance, suppose that a patentholder has sued defendant A in one jurisdiction (e.g., in New York) for infringement of patent X ('Suit A') and defendant B in another jurisdiction (e.g., in California) for infringement of the same patent ('Suit B'). Suppose that Suit A results in an initial judgment against the patentholder on the ground that patent X is invalid and is upheld on appeal in a final judgment. 

It is a well-known principle of patent law in the United States that if a patent is held invalid upon a final judgment with no further allowable appeals, then under the doctrine of issue preclusion (collateral estoppel), the patent cannot be asserted in any other judgment. Similar effects will result whenever courts are bound to recognize the judgment from another court, e.g., under the doctrine of claim preclusion (res judicata), or in the United States, the Full Faith and Credit Clause. In an ideal sense, this negation or voiding of the patent -- and, more generally, the recognition of judgments -- across simultaneously pending ('entangled') lawsuits is instantaneous and thus wholly nonlocal. (We return to this idealization in Section 3.5 below.) 

Further, if the validity of the patent prior to adjudication in Suit A is truly indeterministic, such that,
\begin{align} 
    |\psi_A\rangle = \alpha |\text{Valid}\rangle + \beta |\text{Invalid}\rangle,
\end{align}
then no pre-existing ('real') hidden variable can determine the result of judgment in Suit A and, in turn, the result of judgment in Suit B cannot be determined by a ('real') hidden variable. As such, there is both a nonlocal and contextual (non-real) entanglement between judgment in Suit A and Suit B. Suppose further that the practical effect of a finding of patent validity in Suit A will result in a validity finding in Suit B.\footnote{More realistically, only a decision of invalidity in Suit A will automatically yield a decision of invalidity in Suit B. If patent X is found valid in Suit A, the court in Suit B may still find the patent valid. For simplicity -- and consistent with the practical effect of a validity finding -- we assume otherwise here.}In this case, whatever the judgment in Suit A will in turn dictate the judgment in Suit B. As such, the outcomes are non-separable and must be expressed as follows:
\[
|\Psi\rangle_{AB}
= \alpha \, |{\rm Valid/Valid}\rangle
+ \beta \, |{\rm Invalid/Invalid}\rangle .
\]
Next, assume that validity-invalidity determination is 50/50. In this case:
\[
|\Psi\rangle_{AB} = \frac{1}{\sqrt{2}} \left( |\text{Valid/Valid}\rangle + |\text{Invalid/Invalid}\rangle \right)
\]
Of course, this state is analogous to the maximally entangled Bell state:
\[
|\Phi^+\rangle = \frac{1}{\sqrt{2}} \left( |00\rangle + |11\rangle \right)
\]
Like the maximally entangled Bell states, the judgments associated with the patents are 'non-separable' in the sense that the outcomes of the two states cannot be separated from one another, nor are they classically realistic, because -- at least on the post-classical approach to adjudication -- they are only realised following a measurement (i.e., a judgment). 

We can also model partial entanglement in the adjudicative context. As we noted in the context of interpretive entanglement, external indeterminacy may transform the pure entangled state to a mixed entangled state. For example, suppose Suit A involves patent X and Suit B involves patent X' -- a closely related but different patent. In this situation, a judgment in Suit A may affect any judgment in Suit B, but 'noise' from the difference between patent X and patent X' may introduce external indeterminacy that results in an overall mixed entangled state prior to measurement. 

The discussion of adjudication in this subsection has implicitly assumed that a specific governing rule applied to the dispute at-hand. In some situations, there is no applicable rule, or the court seeks to update a previous rule. In these cases, particularly in the common law tradition, courts will often engage in what is essentially lawmaking, similar to that of legislatures. The sort of entanglement from judges fashioning law is more akin to formulative entanglement, which will typically result in a global updating of 'entangled' states. As discussed earlier, this global updating is less analogous to physical quantum entanglement, because it essentially violates the no-signaling limitation.

\subsection{Entanglement as a Complexity Measure}\label{sec:Advanced Formulations of Legal Entanglement}

In this section, we briefly illustrate a more nuanced example of the type of analysis that can be achieved using our quantum-inspired construction of legal entanglement. We often speak of two quantum systems as being entangled or not, which could lead one to believe that entanglement is a binary or Boolean characteristic, but this is not the case. As noted earlier, systems can exhibit zero, partial, or maximal entanglement. In quantum theory and quantum computing, the extent to which systems are entangled is of great interest, and this has prompted the development of entanglement measures, which quantify the level of entanglement between systems. 

As a benefit of constructing our quantum-like legal systems, we can adapt and apply these measures of entanglement, as well as other quantitative measures, to legal states to quantify the relationships between them. This is useful for a variety of reasons, including legislative complexity analysis and quantifying the impact of amendments and judicial decisions. To illustrate, we perform a more granular analysis of the phrase 'significant detriment' as used in Sections 1023D and 301D of the \textit{Corporations Act} and \textit{NCCPA}, which we previously explored in Section \ref{sec:Interpretive}. 

Sections 1023D and 301D permit the Australian Securities and Investments Commission ('ASIC'), Australia's corporate regulator, to create product intervention orders ('PIOs') regarding financial products and credit products. These PIOs can place significant restrictions on the types of conduct related to these products and, hence, ASIC must be satisfied that certain conditions have been satisfied prior to making a PIO, including the presence or likelihood of 'significant detriment'.

To begin, consider the phrase as it appears in subsections (1) and (3) of Section 1023D. Subsection (1) is specifically concerned with 'significant detriment' that has been, will, or is likely to be caused by a particular financial product, while subsection (3) considers the same risk in the context of \textit{classes} of financial products, rather than a particular product. We might therefore query whether, in our model of Section 1023D, 'significant detriment' ought to be treated as a phrase with a static meaning, or whether the phrase's meaning might vary across subsections (1) and (3)'s differing contexts. 

Per Australia's modern approach to statutory interpretation, where a phrase appears multiple times in a piece of legislation, a consistent meaning should generally be given to that phrase absent contrary intent (\textit{Tabcorp Holdings Limited v Victoria} (2016) 328 ALR 375 at [65]). Thus, we might deem 'significant detriment' to carry the same meaning in both subsections, and in our quantum-like formalism, maximally entangle the two interpretations accordingly. For example, we considered in Section \ref{sec:Interpretive} whether or not non-monetary detriment could suffice for a finding of 'significant detriment'. Regardless of how the court resolves this interpretive question, it is arguable that the answer would be identical in both subsection (1) and subsection (3) circumstances.

Yet, in determining whether a contrary intent exists, the modern approach to statutory interpretation requires us to consider the legislative provision contextually. Given that subsection (1) permits ASIC to make a PIO restricting conduct with respect to a \textit{specific} product, whereas subsection (3) permits the creation of a PIO with respect to an entire \textit{class} of products, one could argue that Parliament did not intend the meaning of 'significant detriment' to be static across both subsections but rather that the threshold of detriment be raised in the context of subsection (3). For example, we also considered Alice's uncertainty regarding whether this detriment ought to be widespread. It is not unreasonable to query whether, given the broader reach of a PIO made under subsection (3), Parliament might have intended for the 'significant detriment' in that subsection to require widespread detriment, while still permitting ASIC to make a PIO regarding a specific financial product under subsection (1) where that product has significantly harmed just a single consumer or a limited pool of specific consumers.

\begin{figure}[ht]
    \centering
    \includegraphics[width=\textwidth]{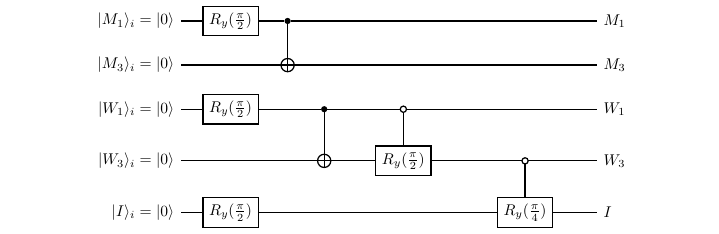}
    \caption{A quantum circuit encodes a potential interpretation of `significant detriment' for the purposes of section 1023D of the \textit{Corporations Act}, taking into account ambiguities regarding a necessary monetary or widespread component of the detriment, and whether an individual product can constitute a class of financial products for the purposes of s 1023D(3).}
    \label{fig:sigDet_circuit}
\end{figure}

We can represent a quantum-like encoding of 'significant detriment' as in Figure \ref{fig:sigDet_circuit}. Here, $M_j$ and $W_j$ are qubits representing a potential requirement of monetary or widespread detriment, respectively, for a finding of 'significant detriment' under subsection ($j$). We also include a qubit labelled $I$, which represents the potential for a single financial product to constitute a class for the purposes of subsection (3).\footnote{For completeness, we note that this latter interpretive ambiguity was in fact resolved by Stewart J of the Federal Court of Australia in \textit{Cigno Pty Ltd v Australian Securities and Investments Commission} [2020] FCA 479. There, it was held that a class could validly comprise a single financial product. This was upheld by the Full Federal Court on appeal in \textit{Cigno Pty Ltd v Australian Securities and Investments Commission} [2021] FCAFC 115 at [71]-[78], though we will artificially treat this legal question as unresolved for pedagogical purposes.} Both pairs, $\{M_1, M_3\}$ and $\{W_1, W_3\}$ are entangled such that, if the subsection (1) component is measured in state $|1\rangle$, so too must the subsection (3) component, representing the 'consistent meaning' principle. However, an anti-control exists from $W_1$ to $W_3$, such that, even if subsection (1) does not require widespread detriment to be satisfied, there is still a chance that subsection (3) does require this element, allowing for a potential deviation from the 'consistent meaning' principle. We also use an anti-control gate from $W_3$ to $I$, representing the notion that if widespread detriment is \textit{not} required to satisfy 'significant detriment' from a class of financial products, it is arguably slightly more likely that an individual product could also constitute a class of products.

Beyond such potential intratextual entanglement within section 1023D, we can also consider \textit{inter}textual entanglement between sections 1023D and 301D. Both sections are almost identical, differing only by reference to \textit{credit} products in s 301D, rather than s 1023D's references to financial products. Hence, we might once again put forth an argument that s 301D could be entangled with s 1023D such that any legal measurement of 'significant detriment' in one section fixes the interpretation of the other section identically. As in the intratextual case above, arguments could be made for or against an intertextual dependency based on the application of statutory interpretation principles. Without preferring one interpretation over the other, we instead introduce an \textit{ancilla} qubit, which we label $\text{Int}$ and which represents whether we should apply an intertextual reading such that 'significant detriment' has the same meaning across sections 301D and 1023D. 

\begin{figure}[ht]
    \centering
    \includegraphics[width=\textwidth]{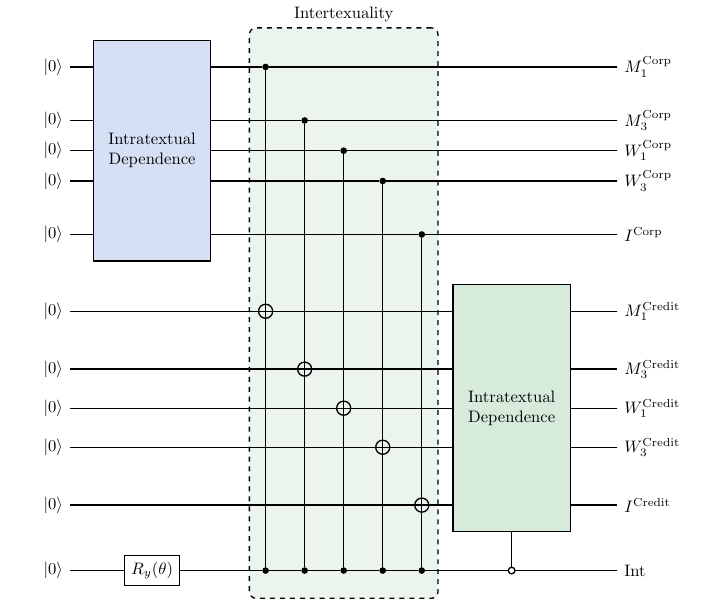}
    \caption{A quantum circuit encodes the intertextual interpretation of `significant detriment' for the purposes of sections 1023D and 301D of the \textit{Corporations Act} and the \textit{NCCPA} respectively. The extent to which we expect an intertextual relationship between the provisions is varied by $\theta$, while the multi-qubit gate, `Intratextual Dependence' represents the circuit given in Figure \ref{fig:sigDet_circuit}.}
    \label{fig:int_circuit}
\end{figure}

Using this ancilla qubit, we can apply the circuit in Figure \ref{fig:int_circuit} to produce our entangled system of qubits representing the phrase 'significant detriment' in sections 301D and 1023D. There, the 'Intratextual Dependence' gate is the five-qubit sub-circuit represented in Figure \ref{fig:sigDet_circuit}. We first apply that gate to the set of qubits representing ambiguities in the \textit{Corporations Act} provision, s 1023D. Then, when Int is in state $|1\rangle$, we simply copy the states of each s 1023D qubit to its corresponding s 301D qubit using controlled gates. Otherwise, if Int is in state $|0\rangle$, we simply apply the same Intratextual Dependence sub-circuit to s 301D's qubits.

\begin{figure}[ht]
    \centering
    \includegraphics[width=\textwidth]{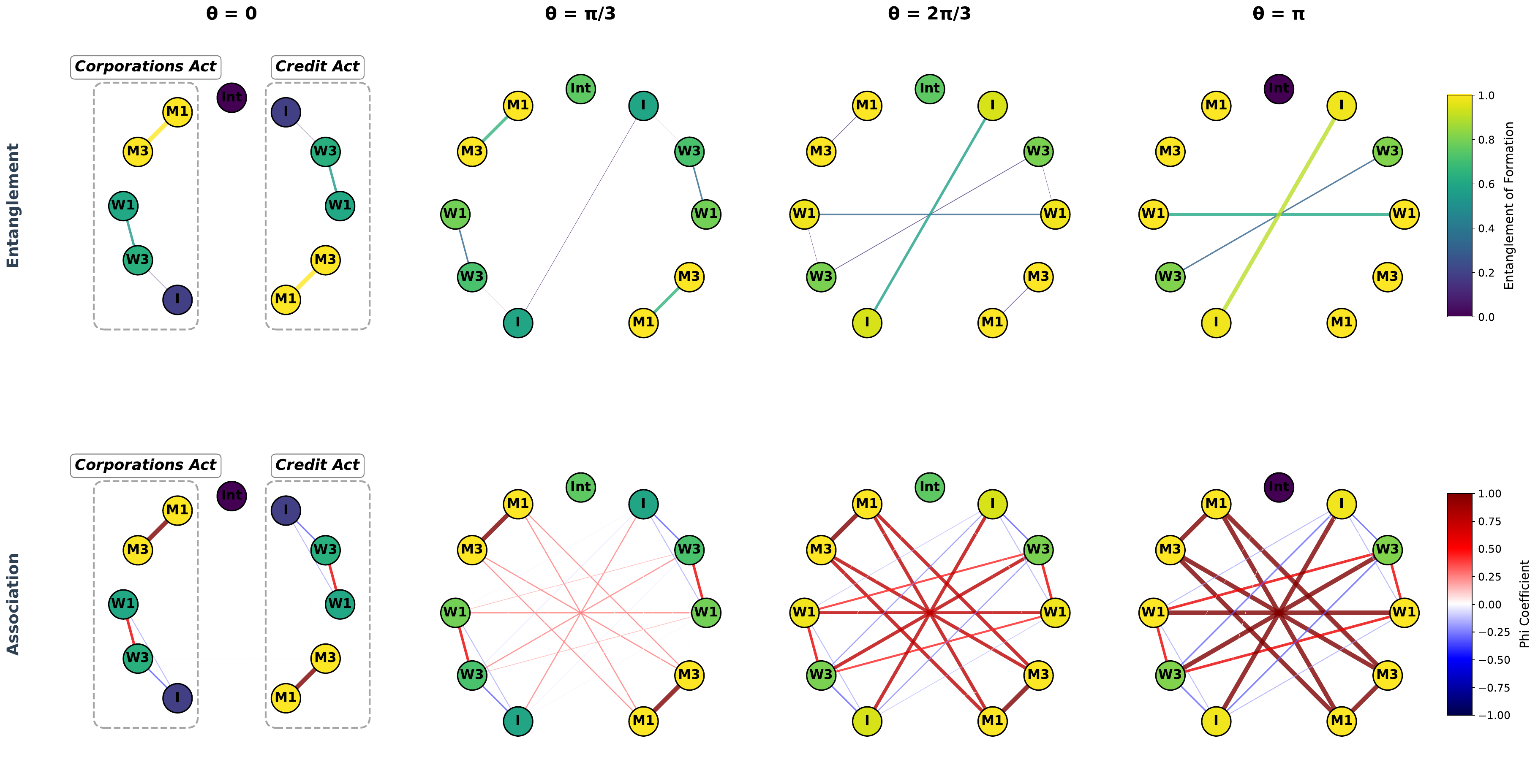}
    \caption{The von Neumann entropy of each qubit in Figure \ref{fig:int_circuit} is plotted as node colours for varying $\theta$ values. Edges between nodes represent the entanglement of formation between pairs of qubits (top) and the association between qubit measurements in the computational basis (bottom).}
    \label{fig:circuit_analysis}
\end{figure}

Once we have constructed our model of these interpretive components, we can analyse the extent to which they are entangled. There are several competing measures of multipartite entanglement, and we seek not to step into that literature here. Instead, we consider bipartite entanglement measures. Specifically, we use the von Neumann entropy (also known as the 'entanglement entropy' in this context) to calculate the extent to which a given interpretive component is entangled with the rest of our model of interpretive components, and we use the concurrence of a two-component subsystem of our model to calculate the entanglement of formation between two interpretive components. We plot these values in Figure \ref{fig:circuit_analysis}, alongside the association of each interpretive component, which represents the classical correlations between those components being held valid or invalid. We vary $\theta$, the angle that represents the state of the Int qubit such that as $\theta$ increases from $0$ to $\pi$, we model a world in which sections 301D and 1023D are increasingly entangled. 

As one might expect, the association between individual interpretive components grows significantly with $\theta$, because as we accept intertextual links between the two Acts, we expect greater correlations between them. However, a deeper insight into our model is gained by considering the entanglement entropy of each interpretive component, and the entanglement of formation between each pair of components. 

When we deny intertextual dependencies between the two Acts ($\theta = 0$), there is entanglement present between subsections (1) and (3) of the two provisions, though no entanglement between the provisions themselves. $M1$ and $M3$, for example, are maximally entangled in both provisions. This entanglement of formation present between two such components essentially measures the extent to which those interpretive components are dependent upon one another when nothing else is known about the state of the broader system. We treat $M1$ and $M3$ as essentially representing the same legal concept -- that 'significant detriment' either does, or does not, require detriment of a monetary nature. Hence, the maximal entanglement present between $M1$ and $M3$ when $\theta=0$ informs us that the correlation between this characteristic of 'significant detriment' in subsections (1) and (3) is not merely a coincidence, nor the result of the application of the same interpretive tests to both subsections. Instead, it is a representation that in this model we cannot distinguish between this characteristic in subsections (1) and (3) -- the monetary nature of 'significant detriment' is one characteristic that can only be considered in the context of both subsections at once. 

As we increase the amplitude of the Int qubit, we see the effects of intertextual relationships between sections 301D and 1023D. The von Neumann entropy of each other individual qubit (represented by that qubit's node colour) increases, representing greater entanglement between individual interpretive components and the rest of the model. However, the entanglement of formation present between pairs of interpretive components in the same Act tends to decrease. At $\theta = \pi$, for example, we find zero entanglement of formation between $M1$ and $M3$ in each provision. This is not to say that there is no relationship between those components anymore. Rather, this is a result of looking only at the entanglement of formation between pairs of qubits. Where $\theta = \pi$, our model suggests that $M1$ and $M3$ of the \textit{Corporations Act} and $M1$ and $M3$ of the \textit{Credit Act} are entangled to the extent that we can no longer describe the monetary characteristic of 'significant detriment' in one provision at a time. Instead, this characteristic can now only be described in the context of both provisions. Indeed, if we calculate the reduced density matrix describing $M1$ and $M3$ across both provisions, we find a GHZ-like maximally entangled 4-qubit state.

\subsection{Practical and Theoretical Implications}\label{sec:Implications}

Several practical and theoretical implications follow from our treatment of legal entanglement. First, we have offered a means to formally model how the interpretation and adjudication of one set of legal relations or laws can immediately and automatically update another set of 'entangled' legal relations or laws. Similarly, the formulation of the law typically is a 'second-order' Hohfeldian, global process that immediately updates all 'entangled' legal relations, and must be described by non-unitary operators. 

An immediate criticism is that our model is the proverbial 'sledgehammer cracking a walnut'. Why not simply use propositional logic -- e.g., 'If Court A decides X, then Court B must decide X' -- to model entanglement instead? We offer three responses. One, as we noted, entanglement in the law is not always a binary, either/or proposition. Rather, in many situations, an interpretation or adjudication of a law or legal relation will only be partially entangled with another law or legal relation, making traditional propositional logic inapt as a model. Although one could turn to fuzzy logic to model such a relation, for reasons we have both discussed at length elsewhere (\cite{sichelman2024}Sichelman, 2025a; \cite{godfrey2024}Godfrey, 2024), it is much more straightforward to use a quantum model both as a matter of practicality and principle. In short, the quantum approach is much more developed than fuzzy logic and is also more consistent with post-classical legal theory's preference for ontological indeterminacy. Two, even when there is maximal entanglement, to the extent courts are truly eliminating ontological indeterminacy, a quantum model more accurately describes the nature of adjudication and interpretation than classical logic, which is deterministic in nature. Three, our model naturally extends to the formulation of the law, which given its effect on numerous relations simultaneously, would be difficult to model using fuzzy logic or a classical approach.

A second implication of our formal model is that it offers a practical means to construct a legal AI system that maintains records of entanglement among laws and legal relations, both statically and dynamically. Akin to the matrix formulation of quantum mechanics, our approach allows legal relations involving specific actors and actions to be represented as qubits of larger systems of mixed and pure states that dynamically evolve in the face of legal interpretation, adjudication, and formulation. 

For instance, following the approach of  \cite{SichelmanSmith2024Network} Sichelman \& Smith (2024), legal actors can be modeled as nodes in a network with edges representing the legal relations between them, such as rights, duties, powers, and the like. Such a network will evolve dynamically based on 'external' changes in ordinary facts as well as 'internal' changes in the law from legislatures, courts, and administrative bodies. Following the treatment here, edges that represent the entanglement of different legal relations may be layered onto the network, so that when one legal relation changes, the entangled relation changes suitably. Attributes of the entanglement edge, such as its thickness, may indicate the extent of the entanglement, such as whether it is maximal or partial. 

If two entangled legal states are pure bipartite states (as opposed to mixed states), one can use the entanglement entropy as a measure of the thickness of the edge to indicate the extent of entanglement. Other measures may be used for a mixed bipartite state (\cite{horodecki2009} (Horodecki et al., 2009). Following the recipe of \cite{SichelmanSmith2024Network} Sichelman \& Smith (2024), the networks formed by these relations, including the entanglement edges, may be grouped into communities, which allows for a quantitative measure of the 'modularity' of legal subsystems. The more modular the system, in general, the lower the 'information costs' related to analyzing the legal relations and related entanglement of the system. Thus, our method provides -- at least at a conceptual level -- a recipe for quantifying the level of modularity and information costs in entangled legal systems.

Finally, to the extent physical entanglement operates in a similar fashion to legal entanglement, our model helps to explain the instantaneous and seemingly non-local character of physical entanglement. Returning to Bell's remark that 'When the Queen dies in London . . . the Prince of Wales, lecturing on modern architecture in Australia, becomes \textit{instantaneously} King' \cite{bell2004} (Bell, Speakable and Unspeakable, 2nd ed., 2004 p. 234), we can imagine an idealization of law in which acts with legal consequence ('legal acts') instantaneously update entangled legal states. 

In actuality, when the Queen dies in London, when a court issues a decision, or when a legislature enacts a law, there are well-known commercial and governmental legal databases (such as Lexis-Nexis, Westlaw, vLex, PACER) that document these legal acts, which are then read and interpreted by legal actors who implement and act on the documented information. Yet, as did Bell, one can imagine that these legal databases operate instantaneously, so that when the Queen dies in London, there is a local update of the applicable 'informational space' that instantaneously propagates to Australia.  

If the legal informational space were separate from the underlying physical reality, so that it was not bound by ordinary physical laws, yet globally connected to every portion of the physical reality, it could act as an external communication channel to enable effective instantaneous updating of legal relations (cf. \cite{sichelman2025} Sichelman, 2025b). Such updating processes are typically 'second-order' in the sense that they concern changes in underlying 'first-order' laws and rules that govern human behaviour. In simpler terms, imagine that a database such as Lexis-Nexis instantaneously updates and is instantly accessible whenever a legal act occurs.

If physical quantum entanglement operates similarly to this idealization, one can posit an analogous 'physical' informational space that acts as an external communication channel to modulate and control events in the underlying physical reality to which it is globally connected \cite{sichelman2025} (Sichelman, 2025b). More specifically, the informational space would catalog and monitor all physical systems, including the relations among those systems, so that when System A is measured and is entangled with System B, the possibilities of measurement for System B will be instantaneously constrained. In essence, the mechanism behind quantum entanglement on this approach is analogous to higher-order legal rules, such as claim and issue preclusion, that mandate when decisions by one court bind decisions by another court. Notably, following the legal model, such an approach to physical quantum entanglement is both nonlocal and classically non-real (i.e., not explainable by classical hidden variables), because, as noted earlier, the outcome of a legal decision is ontologically indeterminate prior to judgment. 

Of course, one may retort that the physical and legal worlds are quite different animals and that although the analogy may have some relevance to modeling entanglement in the law, applying it in the reverse goes too far. More specifically, an external communication channel that could instantly send signals between far-flung parts of the universe would seemingly violate the 'no-signaling principle', which is enshrined in physics by the special theory of relativity. Second, our approach requires an additional quasi-metaphysical, informational space that seemingly cannot be directly accessed. 

Such concerns are all valid on their face, and we can only sketch replies here. First, if we seek to actually explain quantum entanglement through some causal mechanism whereby a measurement in region A affects a measurement in a remote, spacelike-separated region B, that mechanism must be nonlocal in nature. In our legal-inspired mechanism, the no-signaling principle remains fully intact in physical space per se. Rather, it is only in the 'informational space' that modulates or controls the physical space -- namely, the external communication channel -- that a signal may propagate faster than light.\footnote{Because our information channel is not embedded in physical space, but instead in an external and essentially inaccessible informational space, it notably differs from the proposal of (\cite{eberhard1978}Eberhard (1978).}And such signals are confined to measurement events, not ordinary state evolution. Because the measurement events are -- at least from the perspective of an observer in physical space -- entirely random, there is no possibility of faster-than-light signaling within the physical space. 

Second, granted that such an informational space adds a new -- and seemingly inaccessible -- ontological layer to an already complex model of the physical universe. On the other hand, quantum mechanics -- including quantum field theory and the standard model more generally -- arguably comprise inaccessible, ontological layers appended to physical theories. Specifically, the only aspect of quantum mechanics that is accessible is the result of a measurement. State vectors, quantum fields, unitary evolution, and the like are all theoretical models that result in correct predictions of experiments, but are not directly accessible by any observation. Indeed, there is no agreement on whether the quantum state vector is ontological or epistemic in nature -- and, even if it is ontological, exactly how it is constituted. To the extent an informational space assists with model development and predicting the outcome of experiments, then presumably it is just as scientific as other \textit{indirectly} measurable aspects of physical theories. 

Ultimately, the utility of such an approach must be explored beyond the bounds of this article. Our attempt here is merely to indicate that such exploration may indeed be fruitful for theoretical and practical reasons as an explanatory framework for entanglement.

\section{Conclusion and Future Research}\label{sec4}

Previous scholars have recognized that physical entanglement can be used to metaphorically describe certain aspects of the law. Here, we offer not only a more robust conceptual account of how quantum entanglement may be applied to legal systems but also a detailed quantitative formalism that lends itself to notable practical and theoretical endeavors. For instance, the formalism can be used to model interpretive, formulative, and adjudicative entanglement among legal relations and help to explain the role of information costs and 'modularity' in the law and legal artificial intelligence. Finally, if one assumes that the underpinnings of quantum entanglement are similar to those of legal entanglement, one can apply our model in reverse to better understand the nature of physical entanglement. Specifically, on this approach, quantum entanglement would turn on 'second-order' physical processes that are informational in nature and akin to the 'second-order' legal processes underlying how judgments of one court may bind another. Such processes are non-local but also non-realistic, in the sense that no ordinary hidden variable could fully describe the outcome of measurement. 

We hope that these findings lay the groundwork for further work in this fertile intersection of law and physics. First, our approach here is primarily conceptual, illustrated by hypotheticals. A valuable practical application of our approach -- which would also lead to extensions and refinements -- would be to generate entanglement measures using actual datasets. For instance, referring back to the patent validity example, using datasets of related patents, how can legal entanglement be used to model the effect of decisions in one case on decisions in other cases? Alternatively, using actual legal texts and cases interpreting those texts, how can entanglement measures be used to quantify the level of entanglement among interpretive decisions? And so forth. Second, we proposed that entanglement entropy could be used as a measure to model the level of modularity within legal systems, and the role that modularity plays in reducing information costs in determining and adjudicating legal relations among networks of legal actors. Again, how would such a model be constructed using real-world data? For example, in a dataset of M \& A (merger and acquisition) agreements among networks of acquirers and acquired companies, how would the interpretation in a provision in a network cascade throughout the network given its underlying entanglement entropy? These and similar questions require significant empirical as well as modeling efforts, but we hope that our initial formal treatment of legal entanglement will provide an initial path to answer these and other important questions in the burgeoning field of social physics as it applies to law.

\backmatter

\bmhead{Acknowledgements}

pending

\section*{Declarations}

\begin{itemize}
\item Funding: None.
\item Conflict of interest/Competing interests: There are no conflicts of interest.
\item Ethics approval and consent to participate: Not applicable.
\item Consent for publication: Not applicable.
\item Data availability: Not applicable.
\item Materials availability: Not applicable.
\item Code availability: Not applicable.
\item Author contribution: Both authors contributed equally to the manuscript. 
\end{itemize}

\noindent

\bigskip
\begin{flushleft}%

\begin{appendices}




\end{appendices}


\bibliography{sn-bibliography}

\end{flushleft}
\end{document}